\documentclass[a4paper,11pt]{article}
\usepackage{pos}
\usepackage[english]{babel}
\usepackage[utf8]{inputenc}
\usepackage[T1]{fontenc}
\usepackage{pifont}
\usepackage{hyperref}
\usepackage{color,hyperref}
\usepackage{cleveref}
\usepackage{amsmath} 
\usepackage{graphicx}
\usepackage{booktabs}
\usepackage{multirow}
\usepackage{subfigure}
\usepackage{float}
\usepackage{bm}
\newcommand{\beq}{\begin{eqnarray}}
\newcommand{\eeq}{\end{eqnarray}}
\newcommand{\be}{\begin{equation}}
\newcommand{\ee}{\end{equation}}
\newcommand{\bit}{\begin{itemize}}
\newcommand{\eit}{\end{itemize}}

\newcommand{\rw}{\rightarrow}

\title{A new method for a lattice calculation of $\eta_c \rw 2\gamma$}

\author*[a,b]{Yu Meng}

\affiliation[a]{School of Physics,Peking University,
 Beijing 100871,China}

\affiliation[b]{Center for High Energy Physics, Peking University, 
 Beijing 100871, China}
 
\emailAdd{mengyu@pku.edu.cn}

\abstract{We propose a direct method to calculate the decay width of a hadron decaying to two photons, 
which is conventionally obtained by an extrapolation for various off-shell form factors.
The new method provides an simple way to examine the finite-volune effects.
As an example, we apply the method to $\eta_c\rw 2\gamma$. 
Using three $N_f=2$ twisted-mass gauge ensembles with different lattice spacings,
we obtain the final decay width  $\Gamma_{\eta_c\gamma\gamma}=6.57(15)_{\textrm{stat}}(8)_{\textrm{syst}}$ keV, where the systematic error accounts for the fine-tuning of the charm quark mass.
This method can also be applied for other processes which involve the leptonic or radiative particles in the final states.}

\FullConference{%
 The 38th International Symposium on Lattice Field Theory, LATTICE2021
  26th-30th July, 2021
  Zoom/Gather@Massachusetts Institute of Technology
}

\begin{document}
\maketitle

\section{Introduction}
Charmonium decay is a crucial object in particle physics because of its multiscale features~\cite{Bram:2011}.
It presents an ideal laboratory for connecting the perturbative and nonperturbative QCD, and 
tests the validity limit of these various approximations and techniques which work on both sides.
In the nonrelativistic limit, the two-photon decay width of charmonium is directly related to the wave function at the origin~\cite{Bodwin:1994jh}, which acts as a fundamental role in the evaluation of  
charmonium decay and spectrum. The lowest charmonium decay
$\eta_c \rw 2\gamma$ has 
attracted extensive attentions from both experiments \cite{CLEO:2008,BES:2013} and theories
\cite{Dudek:2006,CLQCD:2016,Chen:2016,Feng:2017,CLQCD:2020,Chuan:2020}. 
  
For the lattice calculation on the decay width of $\eta_c \rw 2\gamma$, the traditional approach is to calculate various off-shell form factors~\cite{Dudek:2006} or amplitude squares~\cite{Chuan:2020} with specific photon virtualities. 
The physical result is finally obtained in a continuous extrapolation with photon virtualities approaching to zero. Inevitably, it leads to model-dependent systematic errors. Besides, it also needs more 
computational resources since many off-shell form factors with different photon momenta are required.
In Ref.~\cite{YuMeng:2021}, we have proposed a method to compute the on-shell form factor
straightforwardly without calculating these off-shell form factors. This method could 
significantly reduce the computational costs, especially for a lattice calculation aimming for percent precision.
 In this talk, we give a brief introduction to the method.

\section{Methodology}
In framework of lattice QCD, the on-shell amplitude of a hadron decaying to two-photon 
 is related to an infinite-volume hadronic tensor  $\mathcal{F}_{\mu\nu}(p)$, which is given by
\be
\mathcal{F}_{\mu\nu}(p)=\int dt\,e^{m_Ht/2}\int d^3 \vec{x}\,
e^{-i\vec{p}\cdot \vec{x}}\mathcal{H}_{\mu\nu}(t,\vec{x}),
\ee
where the hadronic function $\mathcal{H}_{\mu\nu}(t,\vec{x})$ is defined as
\be
\mathcal{H}_{\mu\nu}(t,\vec{x})=\langle 0|\textrm{T}[J_{\mu}^{em}(x)J_{\nu}^{em}(0)]|H(k)\rangle ,
\ee
here we denote the initial hadronic state as $|H(k)\rangle$ with the momentum $k=(im_{H},\vec{0})$. 
The electromagnetic current is chosen as
$J_\mu^{em}=\sum_qe_q\,\bar{q}\gamma_\mu q$ ($e_q=2/3,-1/3,-1/3,2/3$ for $q=u,d,s,c$).
One of the photon momenta is indicated as $p=(im_H/2,\vec{p})$ with $|\vec{p}|=m_H/2$, 
which satisfies the on-shell condition. 
If the  initial state is a pseudo-scalar particle, the hadronic tensor $\mathcal{F}_{\mu\nu}(p)$ 
can be parameterized as
\be
\mathcal{F}_{\mu\nu}(p)=\epsilon_{\mu\nu\alpha \beta}p_{\alpha}k_\beta F_{H\gamma\gamma}.
\ee
By multiplying $\epsilon_{\mu\nu\alpha \beta}p_{\alpha}k_\beta$ to both sides, the on-shell form factor is
extracted  through
\be
F_{H\gamma\gamma}=-\frac{1}{2m_{H}|\vec{p}|^2}\int d^4x\,e^{-ipx}
\epsilon_{\mu\nu\alpha0}
\frac{\partial \mathcal{H}_{\mu\nu}(t,\vec{x}) }{\partial x_{\alpha}}.
\ee
After averaging over the spatial direction for $\vec{p}$, $F_{H\gamma\gamma}$ would be
obtained by
\be\label{eq:F_L}
F_{H\gamma\gamma}=-\frac{1}{2m_{H}}\int d^4x\, e^{\frac{m_{H}}{2}t}\,
\frac{j_1(|\vec{p}||\vec{x}|)}{|\vec{p}||\vec{x}|}\,
\epsilon_{\mu\nu\alpha0}x_{\alpha}\mathcal{H}_{\mu\nu}(t,\vec{x}),
\ee
where $j_n(x)$ are the spherical Bessel functions. The decay width is then given as
\be\label{eq:decay_width}
\Gamma_{H\gamma\gamma}=\alpha^2\frac{\pi}{4}m_{H}^3
F_{H\gamma\gamma}^2
\ee  

More recently, the infinite-volume reconstruction method has been proposed to reconstruct the infinite-volume hadronic function from the finite-volume ones
\cite{XuFeng:2019,Tuo:2019,XuFeng:2020,XuFeng1:2020,Christ:2020,Christ:2021,Ma:2021,Tuo:2021,XuFeng:2021}.
In this work, we apply it for $\eta_c\rw 2\gamma$ decay. 
For the time integral in Eq.~(\ref{eq:F_L}), 
it is natural to introduce an integral truncation $t_s$, leading to the contribution of $F_{H\gamma\gamma}(t_s)$. 
For sufficiently large region $t>t_s$, the hadronic function $\mathcal{H}_{\mu\nu}(t,\vec{x})$ 
is dominated by the $J/\psi$ state if only connected diagrams included.  
Such part of contribution is then calculated as
\beq\label{eq:F_INF}
\delta F_{\eta_c\gamma\gamma}(t_s)= -\frac{1}{2m_{\eta_c}}\frac{e^{|\vec{p}|t_s}}
{\sqrt{m_{J/\psi}^2+|\vec{p}|^2}-|\vec{p}|} \times\int d^3\vec{x}\,
\frac{j_1(|\vec{p}||\vec{x}|)}{|\vec{p}||	\vec{x}|}
 \epsilon_{\mu\nu\alpha0}x_{\alpha}\mathcal{H}_{\mu\nu}(t_s,\vec{x}).
\eeq 
Finally, we combine these two parts together to obtain the total contribution
\be
F_{\eta_c\gamma\gamma}=F_{\eta_c\gamma\gamma}(t_s)+\delta F_{\eta_c\gamma\gamma}(t_s),
\ee 
In the simulation, the hadronic function in the small time region $t<t_s$ is calculated in a finite volume, 
as denoted by $\mathcal{H}^{L}_{\mu\nu}(t,\vec{x})$. The infinite-volume $\mathcal{H}_{\mu\nu}(t,\vec{x})$
is replaced by $\mathcal{H}^{L}_{\mu\nu}(t,\vec{x})$.  Such replacement works well because
i) the difference of $\mathcal{H}^{L}_{\mu\nu}(t,\vec{x})$ and 
$\mathcal{H}_{\mu\nu}(t,\vec{x})$ is caused by the exponentially supressed finite-volume effects.
ii)such finite-volume effects are easy to be verified numerically by introducting a spatial integral truncation $R$ 
and examining the $R$-dependence of $F_{H\gamma\gamma}$.

\section{Lattice setup}
The simulation is performed using three $N_f=2$ flavour twisted mass gauge field ensembles generated by 
the Extended Twisted Mass Collaboration (ETMC)~\cite{ETM:2007} with lattice spacing $a \simeq 0.0667,0.085,0.098$ fm. 
The ensemble parameters are reported in Table.~\ref{table:cfgs}.
We tune the charm quark mass by setting the lattice result of 1) $\eta_c$ mass to its physical value and 2) $J/\psi$ mass to its physical value, respectively, 
and regard the deviation in both cases as systematic error in our analysis. For simplicity, we add the suffix ``-I'' and ``-II'' to the ensemble name to specify the cases of 1) and 2). 

\begin{table}[!h]
\caption{\label{table:cfgs}%
 Parameters of gauge ensembles. From left to right, we list the lattice spacing $a$, pion mass 
 $m_{\pi}$, lattice volume=spatial$^3\times$temporal, light twisted mass $a\mu_{l}$, the number of configurations 
 and timeslice $N_{\textrm{conf}} \times T_s$ and a series of time separation $\Delta t$ used for extracting the ground-state contribution.
  All the $\Delta t$ lies in the range of $0.7-1.6$ fm.}
  \center
\begin{tabular}{ccccccccc}
\hline
\textrm{Ensemble} & $a$(fm) &$m_{\pi}(\textrm{MeV})$ & $L^3\times T$  & $a\mu_{l}$  & $N_{\textrm{conf}}\times T_s$ 
& $\Delta t/a$  \\
\hline
a98           &  0.098    & 365 & $24^3\times 48$ & 0.006      & $236\times 48$   &7:17    \\
a85           &  0.085    & 315 & $24^3\times 48$ & 0.004      & $200\times 48$  & 8:18      \\
a67          &  0.0667   & 300 & $32^3\times 64$ & 0.003     & $197\times 64$   &10:24     \\
\hline
\end{tabular}
\end{table}

 The hadronic function $\mathcal{H}_{\mu\nu}$ is calculated by three-point correlation function 
$\Gamma^{(3)}_{\mu\nu}(x,y,t_i)=\langle J^{em}_{\mu}(x)J^{em}_{\nu}(y)\mathcal{O}_{\eta_c}^{\dagger}(t_i)\rangle$
with $t_i=\textrm{min}\{t_x,t_y\}-\Delta t$. Specifically, it is given as
\be
\mathcal{H}_{\mu\nu}(t_x-t_y,\vec{x}-\vec{y})=\Gamma^{(3)}_{\mu\nu}(x,y,t_i)/[(Z_0/2m_{\eta_c})e^{-m_{\eta_c}(t_y-t_i)}]. 
\ee
where $Z_i=\frac{1}{\sqrt{V}}\langle i|\mathcal{O}_{\eta_c}^\dagger|0\rangle$($i=0,1$) and $V=L^3$. $Z_0$ and $m_{\eta_c}$ are obtained by fitting the two-point function $\Gamma_{\eta_c\eta_c}^{(2)}(t)=\langle \mathcal{O}_{\eta_c}(t) \mathcal{O}_{\eta_c}^{\dagger}(0)\rangle$ with a two-state form
\be
\Gamma_{\eta_c\eta_c}^{(2)}(t)=V \sum_{i=0,1}\frac{Z_i^2}{2E_i} \left(e^{-E_it}+e^{-E_i(T-t)}\right)
\ee
where $E_1, E_0=m_{\eta_c}$ are the first- and ground-state energies of $\eta_c$ meson. The $\eta_c$ propagator is solved by using a stochastic $Z_4$ noise. We also apply  the gauss smearing~\cite{Gauss:1990} 
for the $\eta_c$ quark field and APE smearing~\cite{APE:1987} for the gauge field to reduce the excited-state effects. 
It is found that the excited-state contamination is significant in the region $\Delta t \lesssim 1.6$ fm when the precison reaches 1-3\%. 
Such excited-state contribution can be taken into account by choosing a series of different $\Delta t$ and the final form factor is extracted by a multi-state fit
\be
\label{eq:th_fit}
F_{\eta_c\gamma\gamma}(\Delta t)=F_{\eta_c\gamma\gamma}+\xi\, e^{-(E_1-E_0)\Delta t},
\ee
where $ F_{\eta_c\gamma\gamma}$ and $\xi$ are two undeterminted parameters.
We use local current $J^{em}_{\mu}(x)=Z_Ve_c\bar{c}\gamma_{\mu}c(x)$ as the photon interpolating operator.
The vector renormalization facotor $Z_V$ is estimated by a modified ratio of the two-point function over 
 three-point function
 \be\label{eq:zv_ratio}
 \frac{\Gamma^{(2)}_{\eta_c\eta_c}(t)}{\Gamma_{\eta_c\gamma_0\eta_c}^{(3)}
 (t)}= Z_V\cdot \left[1+e^{-m_{\eta_c}(T-2t)} \right]
 \ee 
where
$\Gamma_{\eta_c\gamma_0\eta_c}^{(3)}=\sum_{\vec{x}}\langle \mathcal{O}_{\eta_c}(t) J^{(c)}_{0}(t/2,\vec{x}) \mathcal{O}_{\eta_c}^\dagger(0)\rangle$ denotes the three-point function with a local operator $J^{(c)}_{0}(x)=\bar{c}\gamma^0 c(x)$ inserted between the initial and final $\eta_c$ state with zero momentum.
Compared with the conventional ratio~\cite{Dudek:2006-2}, the modified one have accounted for the 
the around-of-world effect in $\Gamma^{(2)}_{\eta_c\eta_c}$, 
which leads to a distinct systematic effect at $t\simeq T/2$. 

\section{Lattice results}

\begin{figure}[!h]
	\centering
		\subfigure{\includegraphics[width=0.45\textwidth]{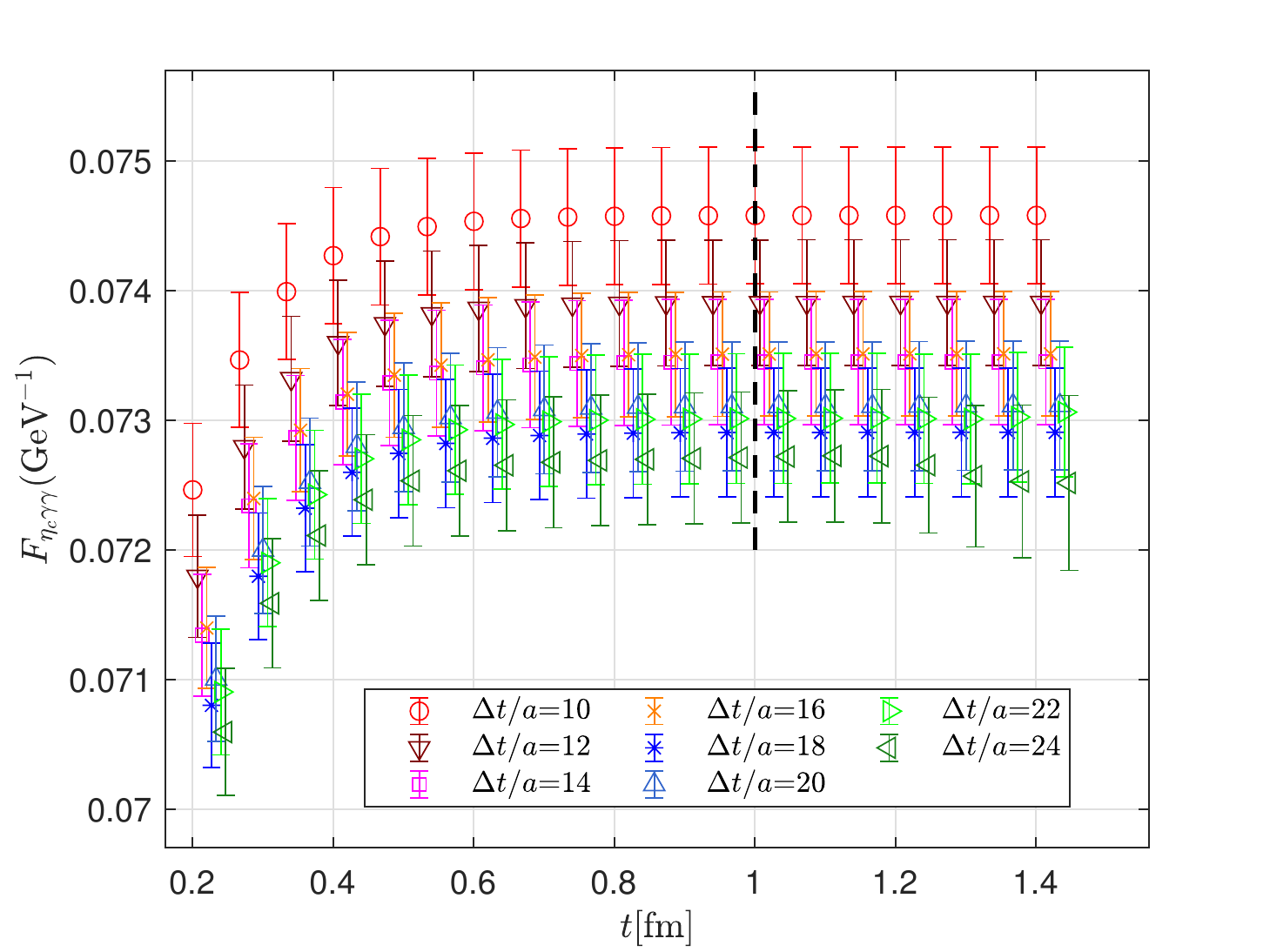}}\hspace{5mm}
		\subfigure{\includegraphics[width=0.45\textwidth]{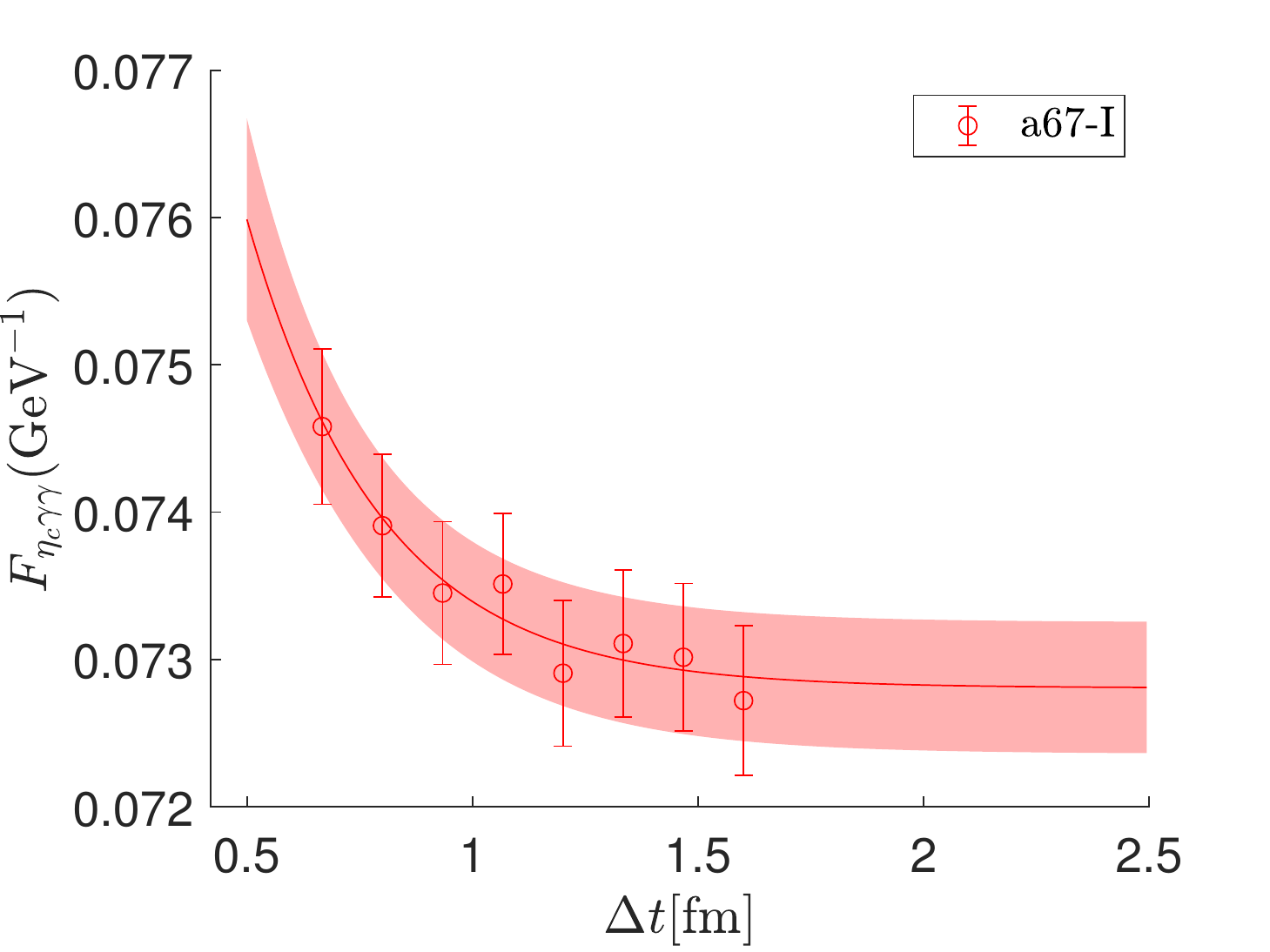}}\hspace{5mm}
    \caption{For ensemble a67-I:
    		In the left panel, lattice results of $F_{\eta_c\gamma\gamma}$ as a function of the 
    		integral truncation $t_s$ with different separation $\Delta t$. The dashed black line
    		denotes the temporal cut $t_s \simeq 1$ $\textrm{fm}$;
    		In the right panel, the ground-state extrapolation for the $F_{\eta_c\gamma\gamma}$.}
    \label{fig:width_a67}
\end{figure}
The lattice results of  $F_{\eta_c\gamma\gamma}$ as a function of the truncation $t_s$ 
with different separations $\Delta t$ are shown in left panel of Figure.~\ref{fig:width_a67}, here we take 
the ensemble a67-I as an example. The integral
in Eq.~(\ref{eq:F_L}) is performed with all $\vec{x}$ summed up and $ t$ 
truncated by $t_s$. We find that a temperal truncation at $t_s\simeq 1$ $\textrm{fm}$ is a conservative choice. 
The results of $F_{\eta_c\gamma\gamma}$ at such truncation are shown in the right pannel.
It shows that $F_{\eta_c\gamma\gamma}$  has an obvious $\Delta t$ dependence,
leading to a significant excited-state effect. 
Finally, we extract the ground-state contribution to $F_{\eta_c\gamma\gamma}$ by a two-state fit
 in Eq.~(\ref{eq:th_fit}).  
 
 \begin{figure}[!h]
 \centering
  \subfigure{\includegraphics[width=0.65\textwidth]{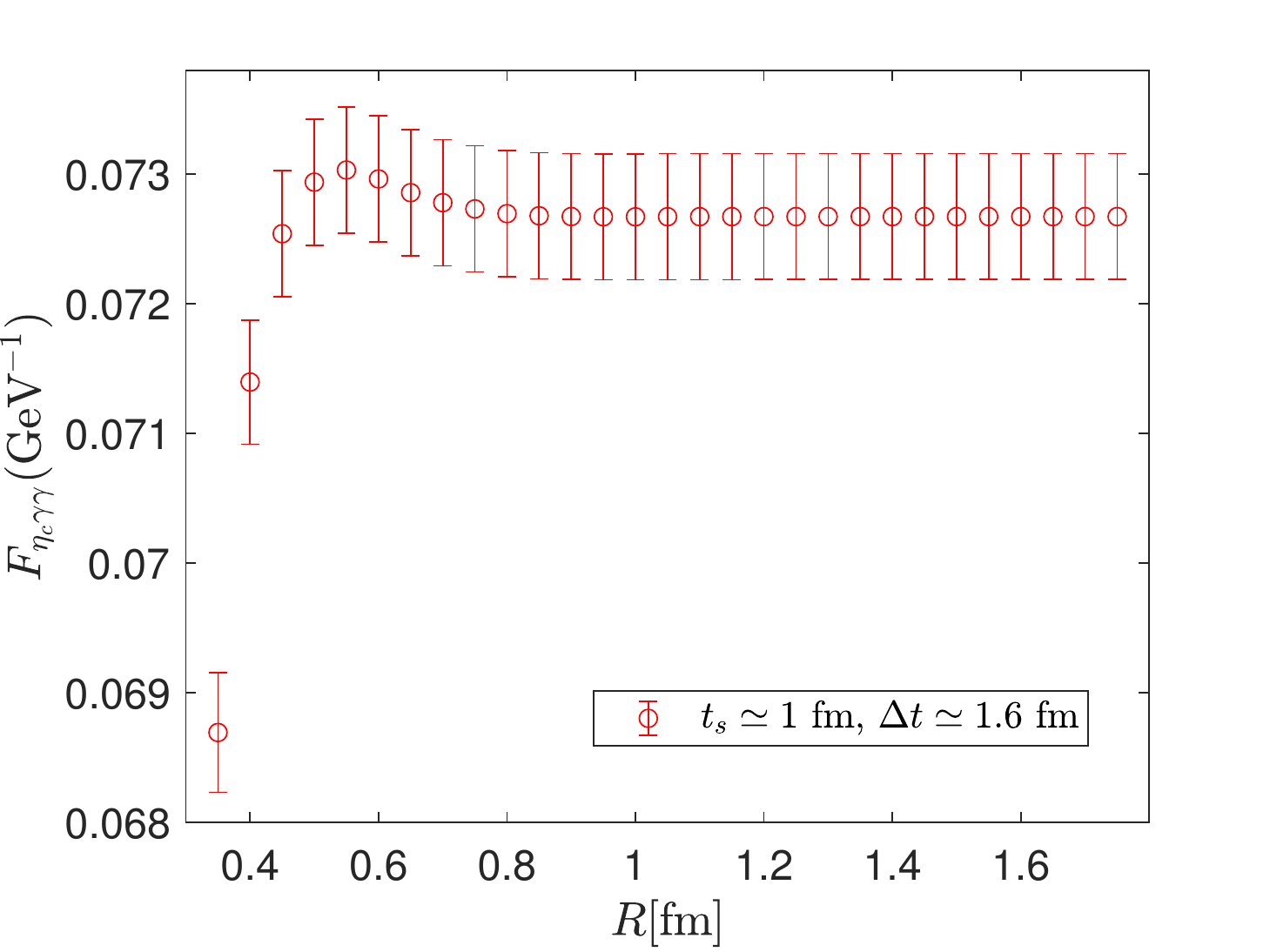}}\hspace{5mm}
    \caption{For ensemble a67-I, $F_{\eta_c\gamma\gamma}$ with $t_s\simeq 1$ fm and $\Delta t \simeq 1.6$ fm as a function of the spatial range truncation $R$.}
    \label{fig:finite}
\end{figure}

The new method allows us to examine the finite-volume effects easily. For the integral in Eqs.~(\ref{eq:F_L}) and (\ref{eq:F_INF}), it is natural to introduce a spatial
integral truncation $R$. The size of the integrand is exponentially suppressed as the $|\vec{x}|$ increases, 
since hadronic function $\mathcal{H}_{\mu\nu}(x)$ is dominated by the $J/\psi$ state at large $|\vec{x}|$.
The form factor $F_{\eta_c\gamma\gamma}$ as a function of $R$ is shown in Figure.~\ref{fig:finite}. 
It is clear that the plateau emerges at almost $R\simeq 0.8$ fm, indicating the hadronic function $\mathcal{H}_{\mu\nu}(x)$
at $|\vec{x}|\gtrsim 0.8$ fm has negligible contribution to $F_{\eta_c\gamma\gamma}$.
All the lattice sizes of ensembles used in this work are greater than  2 fm, which is large enough to accommodate the hadronic system. 
Therefore, the finite-volume effects are well-controlled in our calculation.
 
\begin{figure}[!h]
	\centering
		\subfigure{\includegraphics[width=0.65\textwidth]{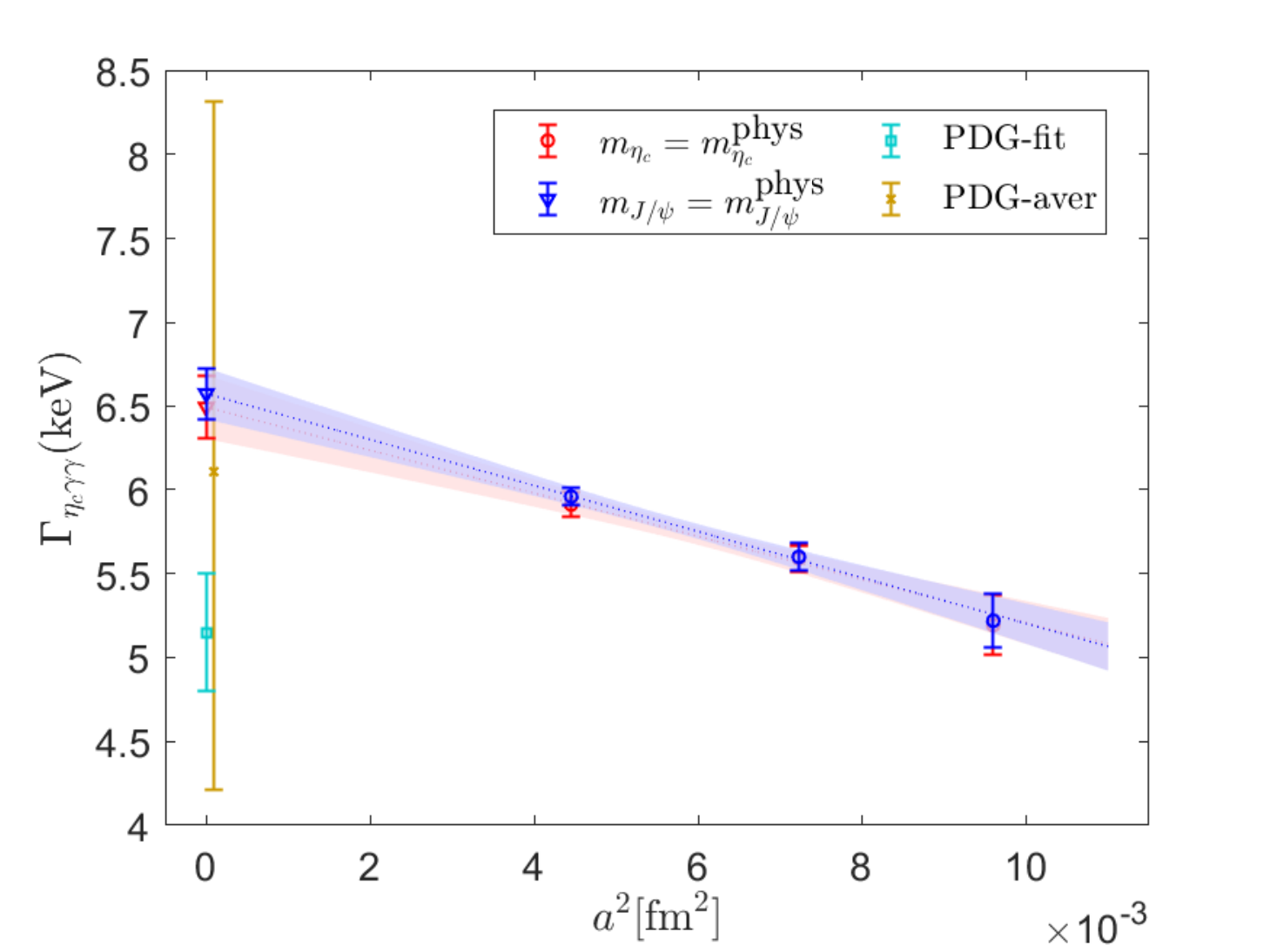}}\hspace{5mm}
    \caption{
    		Decay width $\Gamma_{\eta_c\gamma\gamma}$ under three different lattice spacings. 
    		Experimental results of PDG-fit and PDG-aver are denoted by cyan square and orange cross,
    		respectively. The latter is shifted horizontally for a clear comparison.}
    \label{fig:width_cont}
\end{figure}

The results of decay width $\Gamma_{\eta_c\gamma\gamma}$ are shown in Figure.~\ref{fig:width_cont}. 
A linear extrapolated behavior is well consistent with the lattice values. 
In the continuum limit $a^2\rw 0$, we obtain
\be
\Gamma_{\eta_c\gamma\gamma}=
\left\{
             \begin{array}{lr}
             6.49(19) \, \textrm{keV}  \quad \mu_c \;\textrm{scaled by}\; \eta_c& \\
             6.57(15)  \, \textrm{keV} \quad  \mu_c \; \textrm{scaled by} \; J/\psi   &\\
             \end{array}
\right.
\ee
 where the systematic uncertainty due to fine-tuning of charm quark mass 
is smaller than the statistical error.
We take the value of $m_{J/\psi} \simeq m_{J/\psi}^{\textrm{phys}}$ as the final result and
the deviation as the systematic error. Finally, the decay width is reported as
\be
\Gamma_{\eta_c\gamma\gamma}=6.57(15)_{\mathrm{stat}}(8)_{\mathrm{syst}}\, \textrm{keV},
\ee
For the experimental results, there are two different values quoted by PDG:(i)
 world average of the experiments $B(\eta_c\rw 2\gamma)=1.9^{+0.7}_{-0.6}\times 10^{-4}$;
(ii) combined fit value $B(\eta_c\rw 2\gamma)=(1.61\pm 0.12)\times 10^{-4}$. 
For simplicity, we denote the former as ‘PDG-aver’ and the latter  as ‘PDG-fit’.
Our lattice result is larger than the PDG-fit by 28\% with a 3.6-$\sigma$ deviation, 
but compatible with PDG-aver as it carries a large uncertainty.
The two PDG values are also presented in Figure.~\ref{fig:width_cont} for a comparison.
A recent paper using Dyson-Schwinger equation obtains the decay width as $\Gamma=6.3\sim 6.4$ keV~\cite{Chen:2016}.  Another study from NRQCD including the next-to-next-to-leading-order perturbative correction
gives the branching fraction as $B(\eta_c\rw 2\gamma)=3.1\sim 3.3 \times 10^{-4}$~\cite{Feng:2017},
which is larger than other theoretical  and experimental results. 
For more detailed discussions on these discrepancies, we refer interested readers to Ref.~\cite{YuMeng:2021}.

\section{Conclusion}

In this work we propose a new method to calculate the matrix element of $\eta_c\rw 2\gamma$, 
where the on-shell form factor is obtained straightforwardly by constructing an appropriate 
hadroinc scalar function. It is easy and efficient to examine the finite-volume effects with this method. Besides, it can also be applied for other processes which involve the leptonic 
or radiative particles in the final states, for example, $\pi_0 \rw 2\gamma$~\cite{XFeng:2012}, $J/\psi \rw 3\gamma$~\cite{YM:2020}, $\pi\to e^+e^-$~\cite{NH:lat097}, $K_L\to\mu^+\mu^-$~\cite{NH:lat128}
and radiative leptonic decays $K^-\to\ell^-\bar{\nu}\gamma$, $D_s^+\to \ell^+\nu\gamma$,
$B\to\ell^-\bar{\nu}\gamma$~\cite{CK:2019,AD:2021}. 
In the continuous extrapolation with three lattice spacings, we obtain the decay width $\Gamma(\eta_c\rw 2\gamma)=6.57(15)_{\textrm{stat}}(8)_{\textrm{syst}}$ keV, 
where the systematic effects of excited-state contamination, finite volume and fine-tuning of charm quark mass 
are under well control.

\begin{acknowledgments}
We thank ETM Collaboration for sharing the gauge configurations with us. A particular acknowledgement goes to Carsten Urbach, Chuan Liu and Xu Feng.
We gratefully acknowledge many helpful discussions with Michael Doser, Luchang Jin, Haibo Li and Yan-Qing Ma. 
Y.M. acknowledges support by NSFC of China under Grant No. 12047505 and 
	State Key Laboratory of Nuclear Physics and Technology, Peking University. 
The main calculation was carried out on Tianhe-1A supercomputer at Tianjin National
	Supercomputing Center and partly supported by High-performance Computing Platform 
	of Peking University.
\end{acknowledgments}

\end{document}